\documentclass[10pt,english,aps,showkeys,showpacs,nofootinbib]{revtex4-1}
\usepackage[T1]{fontenc}
\usepackage{textcomp}
\usepackage[utf8]{inputenc}
\usepackage{geometry}
\usepackage{babel}
\usepackage{amsmath}
\usepackage{amssymb}
\usepackage{xcolor}
\usepackage[bookmarks=false, breaklinks=false,pdfborder={0 0 1},backref=section,colorlinks=false]
{hyperref}

\makeatletter

\usepackage{todonotes}
\usepackage[bottom]{footmisc}
\usepackage{graphicx}
\usepackage{dcolumn}
\usepackage{bm}
\usepackage{todonotes}
\usepackage{babel}

\makeatother

\begin{document}

\selectlanguage{english}
\title{Feshbach--Villars oscillator in Kaluza-Klein theory}

\author{Abdelmalek Bouzenada}
\email{abdelmalek.bouzenada@univ-tebessa.dz ; abdelmalekbouzenada@gmail.com}
\affiliation{Laboratory of theoretical and applied Physics, Echahid Cheikh Larbi. Tebessi University, Algeria}

\author{Abdelmalek Boumali}
\email{boumali.abdelmalek@gmail.com}
\affiliation{Laboratory of theoretical and applied Physics, Echahid Cheikh Larbi. Tebessi University, Algeria}

\author{R. L. L. Vitória}
\email{ricardo.vitoria@pq.cnpq.br ; ricardo-luis91@hotmail.com}
\affiliation{Faculdade de Física, Universidade Federal do Pará, Av. Augusto Corrêa,~\\
 Guamá, Belém, PA 66075-110, Brazil}

\author{Faizuddin Ahmed}
\email{faizuddinahmed15@gmail.com ; faizuddin@ustm.ac.in}
\affiliation{Department of Physics, University of Science \& Technology Meghalaya,~~\\
 Ri-Bhoi, Meghalaya 793101, India}

\author{Marwan Al-Raeei}
\email{mhdm-ra@scs-net.org ; mn41@liv.com}
\affiliation{Faculty of Science,Damascus~~\\
 University, Damascus, Syria}

\date{\today}

\begin{abstract}

This research focuses on exploring the relativistic quantum dynamics of spin-0 scalar massive charged particles within the framework of the relativistic Feshbach-Villars oscillator (FVO) in the context of the Kaluza-Klein Theory (KKT). Specifically, we investigate the behavior of these particles in the presence of a cosmic string space-time within the framework of the Kaluza-Klein theory. To begin, we solve the Feshbach-Villars equation in the background of a cosmic string space-time by considering the principles of the Kaluza-Klein theory. We obtain the eigenvalue solution, which provides valuable insights into the behavior of the system. Next, we re-examine this system by considering the Feshbach-Villars quantum oscillator, a specific mathematical representation of the Feshbach-Villars equation. By employing this oscillator, we are able to analytically determine the eigenvalues of the system. This analytical solution allows for a deeper understanding of the system's characteristics. In the subsequent stage, we investigate the interaction between the Feshbach-Villars equation and oscillator in the presence of a cosmic dislocation within the framework of the Kaluza-Klein Theory. We analytically solve the wave equation associated with this system, shedding light on the behavior of the wave function. Furthermore, we analyze the influence of the topological defect (cosmic dislocation) on the quantification of energy and the wave function of the Feshbach-Villars oscillator. We also explore the impact of external fields on the system. These analyses provide valuable insights into how these factors affect the behavior and properties of the Feshbach-Villars oscillator.

\end{abstract}



\maketitle

\section{Introduction }

Gaining insight into how the gravitational field impacts the behavior 
of quantum mechanical systems is a subject that is of interest to
many. One perspective on gravity, as described by Einstein's theory
of general relativity(GR)\citep{key-1}, views it as a geometric feature
of space-time. This theory explains that the classical gravitational
field\citep{key-2} arises from the curvature of space-time. Einstein's
theory of general relativity has been able to accurately predict the
existence of phenomena such as gravitational waves and black holes\citep{key-3}.
In contrast, Quantum Mechanics(QM) is a widely-accepted framework
for understanding the behavior of particles at the microscopic level\citep{key-4},
explaining the interactions between tiny particles and the emergence\citep{key-5}
of three fundamental forces: the weak, strong, and electromagnetic
interactions. However, the development of a unified theory of quantum
gravity, that reconciles general relativity and quantum mechanics,
has been faced with significant obstacles and technical challenges
that have yet to be overcome, at least until recently\citep{key-6,key-7}.

One approach to developing a theory that explores the relationship
between gravity and relativistic quantum mechanics is to extend the
concepts of relativistic particle dynamics in flat Minkowski space
to a curved background geometry\citep{key-8,key-9}. This generalization
can be used to create a comprehensive understanding of how the gravitational
field impacts the behavior of quantum particles\citep{key-10,key-11,key-12},
applicable to a range of models where the notion of curvature appears.
Such an approach can provide predictions for macroscopic observables,
enabling experimental verification of certain phenomenological consequences,
particularly in astrophysics and cosmology. Furthermore, studying
the thermodynamic properties of relativistic particles while accounting
for gravitational effects can yield valuable insights into the fundamental
statistical quantities that underlie the quantum behavior of gravity.

In recent decades, there has been extensive research on topological
defects such as domain walls, cosmic strings, global monopoles, and textures.
These defects are still a highly active field of research study in condensed
matter physics, Cosmology, astrophysics, and elementary particle physics models.
It is believed that these structures arise due to the Kibble mechanism\citep{key-13,key-14,key-15}
during symmetry-breaking phase transitions in the early universe's
cooling process\citep{key-16,key-17}. {\color{red} Among these topological defects}, cosmic strings 
have garnered significant interest (for more information, see\citep{key-18})
due to their ability to produce observable effects such as galaxy
seeding and gravitational lensing. Additionally, studying cosmic strings
can provide valuable insights into the particle physics at very high energies
in various scenarios. Static or rotating cosmic strings can offer
unique opportunities to investigate the properties of these structures
and their implications. Moreover, recent research has suggested that
cosmic strings may exhibit properties similar to superconducting wires,
leading to intriguing consequences in modern physics. {\color{red} In the context of quantum systems, cosmic strings have widely been investigated including external scalar and vector potential in the presence or absence of magnetic and quantum flux fields. These investigations include Landau levels in the presence of disclinations \cite{CF}, dynamics of a charged particle in the presence of magnetic field and scalar potential \citep{ERFM}, spin-0 massive charged particles in four-dimensional curved space-time with cosmic string \citep{FA}, bosonic oscillator field with Cornell potential in cosmic string space-time \citep{MH}, motion of a quantum particle in the spinning cosmic string space-time \citep{HH}, non-central potentials in cosmic string space-time \citep{HH2}, Klein-Gordon field in spinning cosmic-string space-time with the Cornell potential \citep{HH3}, relativistic Landau levels in rotating cosmic string space-time \citep{MSC}, exact solutions of the Klein–Gordon equation in the presence of a dyon, magnetic flux and scalar potential in the spacetime of gravitational defects \citep{ALCO}, Landau levels in the presence of topological defects \citep{GDAM}, relativistic scalar particle with Cornell-type potential in cosmic string space-time with a spacelike dislocation \citep{FA2}, harmonic oscillator in conical singularities \citep{CF2}, bosonic oscillator under external fields in cosmic string space-time with a spacelike dislocation \citep{FA3}, Weyl-fermions in G\"{o}del-type space-time with a topological defect \citep{GQG}, quantum influence of topological defect in G\"{o}del-type space-time \citep{JC}, Dirac oscillator interacting with a topological defect \citep{JC2}, geometric phase for a neutral particle in rotating frames of cosmic string space-time \cite{KB}, Integer quantum Hall effect on an interface with disclinations \citep{AAL}, rotating effects on the Dirac oscillator in cosmic string space-time \citep{KB2}, the Dirac oscillator in a spinning cosmic string spacetime \citep{MH2}, Landau quantization in the spinning cosmic string spacetime \citep{CRM}, PDM Klein-Gordon oscillators in cosmic string spacetime in magnetic and Aharonov–Bohm flux fields \citep{OM}, interaction of the Dirac oscillator with the Aharonov–Casher system in topological defect space-time  \citep{KB3} and many more.}

The harmonic oscillator (HO) has been recognized as an essential tool in theoretical physics for several years \citep{key-19}. As a well-studied, exactly solvable model, it has proven useful for analyzing complex problems within the framework of quantum mechanics\citep{key-20}. The relativistic generalization of the quantum harmonic oscillator has been effective in explaining diverse aspects of molecular, atomic, and nuclear interactions. The complete set of exact analytical solutions available when dealing with such models has led to significantly different explanations of mathematical and physical phenomena. As a result, a wide range of applications can be achieved through the underlying formulation. {\color{red} In quantum mechanical context, the harmonic oscillator problem has been investigated, such as in conical singularities background \citep{CF2}, in an environment with a point-like defect \cite{rllv}  with Mie-type potential in point-like global monopole \citep{PRSA}, and in the background of a topologically charged Ellis-Bronnikov-type wormhole \citep{EPL}.}

The Dirac oscillator (DO) is now recognized as playing a crucial role
in the behavior of several relativistic quantum systems. Itô et al\citep{key-21}.
have pointed out that earlier developments of spin-1/2 particle dynamics
with a linear trajectory showed that the non-relativistic limit of
this system leads to the ordinary harmonic oscillator with a strong
spin-orbit coupling term. The DO itself can be obtained from the free
Dirac equation by introducing an external linear potential through
a minimal substitution of the momentum operator $\hat{p}\longrightarrow\hat{p}-im\omega\beta\hat{r}$,
as described by Moshinsky and Szczepaniak \citep{key-22}. In addition
to the theoretical focus on studying the DO, valuable insights can
be gained by considering the physical interpretation, which is essential
for understanding many relevant applications. {\color{red} This DO has been investigated by many researchers in flat space background with external field, in curved space-time with external fields as well as topological defects background produced by cosmic string (see, Refs. \cite{JC2,KB,AAL,KB2,MH2,KB3} and related references there in).}

The Klein-Gordon oscillator \citep{key-23,key-24} is a quantum mechanical
model that describes a particle that oscillates in a potential well\citep{key-25}
while obeying the relativistic Klein-Gordon equation. The model has
been studied in various contexts\citep{key-26}, including in the
presence of topological defects such as cosmic strings\citep{key-27,key-28}
and domain walls, as well as in the presence of uniform magnetic fields.
One approach to studying the Klein-Gordon oscillator in general background
spacetimes produced by topological defects is through Kaluza-Klein
theory\citep{key-29,key-30,key-31,key-32,key-33,key-34}. This theory
involves the idea of compactifying extra dimensions of spacetime and
has been applied to study the behavior of particles in the presence
of topological defects in condensed matter physics systems. The effects
of topological defects such as cosmic strings, domain walls, and global
monopoles have been shown to play an important role in condensed matter
physics systems \citep{key-35,key-36,key-37,key-38}due to their ability
to compensate for the elastic contribution introduced by the defect
through fine-tuning of the external magnetic field. Recent studies
have investigated the Klein-Gordon oscillator in various spacetimes,
including the Som-Raychaudhuri spacetime in the presence of uniform
magnetic fields. These studies aim to provide a deeper understanding
of the behavior of particles in the presence of topological defects
and other gravitational sources\citep{key-39,key-40,key-41,key-42,key-43,key-44}.
Bouzenada et al\citep{key-45} investigate the Feshbach-Villars oscillator
(FVO) case in spinning cosmic string space-time and discuss some findings(thermal
properties and density of this system). Bouzenada et al\citep{key-46}
investigate the Feshbach-Villars oscillator (FVO) case in cosmic dislocation
space-time under coulomb-potential type.

The structure of this paper is as follows. In the next section, we
derive the FV equations for scalar bosons in Minkowski and static
Cosmic string space-time considering both the free and the interaction
case. We introduce the KG oscillator in a Hamiltonian form, then we
solve the obtained Free equations and FVO in cosmic dispiration background
in a Kaluza--Klein theory, In the next section we solve the obtained
Free equations and FVO in cosmic dislocation in Som--Raychaudhuri
spacetime in Kaluza--Klein theory, and finally a conclusion.We shall
always use natural units $\hbar=c=1$ throughout the article, and
our metric convention is $(+,-,-,-,-)$.

\section{An Overview of the Feshbach-Villars Formalism in flat and curved spacetime}

{\color{red}

\subsection{In flat space-time}

This section discusses the relativistic quantum 
description of a spin-0 particle propagating in Minkowski space-time
using the metric tensor $\eta_{\mu\nu}=\text{diag}\left(1,-1,-1,-1\right)$.The
usual covariant KG equation for a scalar massive particle $\Phi$
with mass $m>0$ is 
\begin{equation}
\left(\eta^{\mu\nu}\,D_{\mu}\,D_{\nu}+m^{2}\right)\Psi (x,t)=0,\label{eq:01}
\end{equation}
The minimally-coupled covariant derivative is denoted by $D_{\mu}=\left(p_{\mu}-i\,e\,A_{\mu}\right)$.
The classical four momentum is $p_{\mu}=\left(E,-p_{i}\right)$, while
the electromagnetic four potential is $A_{\mu}=\left(A_{0},-A_{i}\right)$.
The magnitude of the particle charge is given by e.

It is worth noting at this point that \eqref{eq:01}
may be expressed in Hamiltonian form with the time first derivative,
i.e. as a Schrödinger-type equation \cite{key-47}
\begin{equation}
\mathcal{H}\,\Phi\left(x,t\right)=i\,\frac{\partial}{\partial t}\,\Phi(x,t),\label{eq:02}
\end{equation}
The Hamiltonian $\mathcal{H}$ may be defined using the FV linearization
process, which involves converting \ref{eq:01} to a first order in
time differential equation.

The two-component wave function is introduced,
\begin{equation}
\Phi(x,t)=\left(\begin{array}{c}
\phi_{1}(x,t)\\
\phi_{2}(x,t)
\end{array}\right)=\frac{1}{\sqrt{2}}\left(\begin{array}{c}
1+\frac{i}{m}\mathcal{D}\\
1-\frac{i}{m}\mathcal{D}
\end{array}\right)\psi(x,t)\label{eq:03}
\end{equation}
Here, $\psi(x,t)$ obeys the KG wave equation, and $\mathcal{D}$
is defined in such a way that 
\begin{equation}
\mathcal{D}=\frac{\partial}{\partial t}+i\,e\,A_{0}(x)\label{eq:04}
\end{equation}
The aforementioned transformation \eqref{eq:03} involves inserting
wave functions that meet the requirements. 
\begin{equation}
\psi=\phi_{1}+\phi_{2},\qquad i\mathcal{D}\psi=m\left(\phi_{1}-\phi_{2}\right).\label{eq:05}
\end{equation}
It is more convenient to write for our subsequent review, 
\begin{equation}
\begin{aligned}\phi_{1} & =\frac{1}{2\,m}\left[m+i\,\frac{\partial}{\partial t}-e\,A_{0}\right]\psi\\
\phi_{2} & =\frac{1}{2\,m}\left[m-i\,\frac{\partial}{\partial t}+e\,A_{0}\right]\psi,
\end{aligned}
\label{eq:06}
\end{equation}
Eq. \eqref{eq:01} becomes equivalent 
\begin{equation}
\begin{aligned}\left[i\,\frac{\partial}{\partial t}-e\,A_{0}\right]\left(\phi_{1}+\phi_{2}\right) & =m\,\left(\phi_{1}-\phi_{2}\right)\\
\left[i\,\frac{\partial}{\partial t}-e\,A_{0}\right]\left(\phi_{1}-\phi_{2}\right) & =\left[\frac{\left(p_{i}-eA_{i}\right)^{2}}{m}+m\,\right]\left(\phi_{1}+\phi_{2}\right),
\end{aligned}
\label{eq:07}
\end{equation}
The addition and subtraction of these two equations yield a system
of first-order coupled differential equations 
\begin{equation}
\begin{aligned}\frac{\left(p_{i}-\,A_{i}\right)^{2}}{2\,m}\left(\phi_{1}+\phi_{2}\right)+\left(m+e\,A_{0}\right)\phi_{1} & =i\,\frac{\partial\phi_{1}}{\partial t}\\
\frac{-\left(p_{i}-e\,A_{i}\right)^{2}}{2\,m}\left(\phi_{1}+\phi_{2}\right)-\left(m-e\,A_{0}\right)\phi_{2} & =i\frac{\partial\phi_{2}}{\partial t},
\end{aligned}
\label{eq:08}
\end{equation}
The FV Hamiltonian of a scalar particle in the presence of electromagnetic
the interaction may be expressed using Eqs. \eqref{eq:08} as 
\begin{equation}
\mathcal{H}_{\text{FV}}=\left(\tau_{3}+i\,\tau_{2}\right)\frac{\left(p_{i}-e\,A_{i}\right)^{2}}{2\,m}+m\,\tau_{3}+e\,A_{0}(x),\label{eq:09}
\end{equation}
where $\tau_{i}\,\left(i=1,2,3\right)$ are the conventional $2\times2$
Pauli matrices are given by 
\begin{equation}
\tau_{1}=\left(\begin{array}{cc}
0 & 1\\
1 & 0
\end{array}\right),\quad\tau_{2}=\left(\begin{array}{cc}
0 & -i\\
i & 0
\end{array}\right),\quad\tau_{3}=\left(\begin{array}{cc}
1 & 0\\
0 & -1
\end{array}\right).
\end{equation}

It's worth noting that the Hamiltonian \eqref{eq:09} meets the generalized
hermicity requirement\footnote{\color{red} The Hamiltonian $\mathcal{H}$ is said to be pseudo-Hermitian
If there is an invertible, Hermitian, linear operator $\beta$ such
that $\mathcal{H}^{\dagger}=\beta\mathcal{H}\beta^{-1}$ \cite{Mostafa1, Mostafa2}.}  
\begin{equation}
\mathcal{H}_{\text{FV}}=\tau_{3}\mathcal{H}_{\text{FV}}^{\dagger}\tau_{3},\qquad\mathcal{H}_{\text{FV}}^{\dagger}=\tau_{3}\mathcal{H}_{\text{FV}}\tau_{3}.
\end{equation}
The one-dimensional FV Hamiltonian reduces to for free particle propagation,
i.e., no interaction is assumed left $\left(A_{\mu}=0\right)$. 
\begin{equation}
\mathcal{H}_{0}=\left(\tau_{3}+i\,\tau_{2}\right)\frac{p_{x}^{2}}{2\,m}+m\tau_{3},\label{eq:12}
\end{equation}
The solutions to the time-independent free Hamiltonian are simply
stationary states. Assuming the solution, 
\begin{equation}
\Phi\left(x,t\right)=\Phi\left(x\right)e^{-i\,E\,t}=\left(\begin{array}{c}
\phi_{1}\left(x\right)\\
\phi_{2}\left(x\right)
\end{array}\right)e^{-i\,E\,t},\label{eq:13}
\end{equation}
with $E$ denoting the system's energy. As a result, Eq. \eqref{eq:02}
may be represented as 
\begin{equation}
\mathcal{H}_{0}\,\Phi\left(x\right)=E\,\Phi\left(x\right),\label{eq:14}
\end{equation}
This is the one-dimensional FV equation of the free relativistic spin-0
particle, and it is performed in order to have an alternate Schrödinger-type
to KG equation. Recently, this formalism has been used in the case
of the one-dimensional Dirac oscillator, and the results were in good
agreement with those obtained in the literature.

In what follows, we review the FV formalism in the curved spacetime.

\subsection{In curved space-time}

It is widely known that the generally covariant
relativistic wave equations of a scalar particle in a Riemannian space-time
characterized by the metric tensor $g_{\mu\nu}$ may be found by reformulating
the KG equation so that\citep{key-27} 
\begin{equation}
\left(\square+m^{2}-\mathcal{\xi}R\right)\Phi(x,t)=0,\label{eq:15}
\end{equation}
where $\square$ is the Laplace-Beltrami operator denoted by 
\begin{equation}
\square=g^{\mu\nu}\,D_{\mu}\,D_{\nu}=\frac{1}{\sqrt{-g}}\partial_{\mu}\left(\sqrt{-g}\,g^{\mu\nu}\,\partial_{\nu}\right),\label{eq:16}
\end{equation}
$\xi$ denotes a real dimensionless coupling constant, and $R$ is
the Ricci scalar curvature given by $R=g^{\mu\nu}R_{\mu\nu}$, where
$R_{\mu\nu}$ is the Ricci curvature tensor. The inverse metric tensor
is $g^{\mu\nu}$, and $g=\det\left(g_{\mu\nu}\right)$\citep{key-45,key-46,key-47,key-48,key-49,key-50}.

Now, according the works of Silenko\citep{key-54,key-55},
the case of flat spacetime can be extended to the case of cured space
time as follows: The detail form of (\ref{eq:15}) is
\begin{equation}
\left(\partial_{0}^{2}+\frac{1}{g_{00}\sqrt{-g}}\left\{ \partial_{i}\,\sqrt{-g}\,g^{0i}\right\} \partial_{0}+\frac{1}{g_{00}\,\sqrt{-g}}\,\partial_{i}\,(\sqrt{-g}\,g^{ij}\,\partial_{j})+\frac{m^{2}-\mathcal{\xi}R}{g^{00}}\right)\,\Phi(x,t)=0\label{eq:16.1}
\end{equation}
The anti-commutator is denoted by the curly bracket in Eq. \eqref{eq:16.1}.
Using the following transformation in the components
of the wave function \citep{key-54}
\begin{equation}
\psi=\phi_{1}+\phi_{2},\qquad i\left(\frac{\partial}{\partial t}+\mathcal{Y}\right)\psi=\mathcal{N}\left(\phi_{1}-\phi_{2}\right),\label{eq:16-1}
\end{equation}
where $\mathcal{N}$ is an arbitrary nonzero real parameter and 
\begin{equation}
\mathcal{Y}=\frac{1}{2}\left\{ \partial_{i},\sqrt{-g}\frac{g^{0i}}{g^{00}}\right\} ,\label{eq:16-2}
\end{equation}
we obtain the Hamiltonian form of the KG equation as follows \citep{key-54,key-55}
\begin{equation}
\mathcal{H}_{GFVT}=\tau_{z}\left(\frac{\mathcal{N}^{2}+\mathcal{T}}{2\mathcal{N}}\right)+i\tau_{y}\left(\frac{-\mathcal{N}^{2}+\mathcal{T}}{2\mathcal{N}}\right)-i\mathcal{Y},\label{eq:16-3}
\end{equation}
with\citep{key-54,key-55} 
\begin{align}
\mathcal{T} & =\partial_{i}\frac{G^{ij}}{g^{00}}\partial_{j}+\frac{m^{2}-\xi R}{g^{00}}+\frac{1}{\mathcal{F}}\nabla_{i}\left(\sqrt{-g}G^{ij}\right)\nabla_{j}\left(\frac{1}{\mathcal{F}}\right)+\sqrt{\frac{\sqrt{-g}}{g^{00}}}G^{ij}\nabla_{i}\nabla_{j}\left(\frac{1}{\mathcal{F}}\right)+\frac{1}{4\mathcal{F}^{4}}\left[\nabla_{i}\left(\mathcal{U}^{i}\right)\right]^{2}\nonumber \\
 & \qquad-\frac{1}{2\mathcal{F}^{2}}\nabla_{i}\left(\frac{g^{0i}}{g^{00}}\right)\nabla_{j}\left(\mathcal{U}^{i}\right)-\frac{g^{0i}}{2g^{00}\mathcal{F}^{2}}\nabla_{i}\nabla_{j}\left(\mathcal{U}^{i}\right),\label{eq:16-4}
\end{align}
where
\begin{equation}
G^{ij}=g^{ij}-\frac{g^{0i}g^{0j}}{g^{00}},\qquad\mathcal{F}=\sqrt{g^{00}\sqrt{-g}},\qquad\mathcal{U}^{i}=\sqrt{-g}g^{0i}.\label{eq:16-5}
\end{equation}
We notice that the initial FV transformations are fulfilled for $\mathcal{N}=m$.

}

\section{ THE FV OSCILLATOR OF SPIN-0 PARTICLE IN COSMIC KALUZA-KLEIN THEORY}

\subsection{Free Feshbach-Villars equation in the background of a cosmic string
space-time in Kaluza--Klein theory}

In this part, we will discuss the topological defect that serves as
the foundation for our study. Inspired by the description of an edge
dislocation in crystalline solids, we build a generalization of this
topological defect in gravity. We can observe that an edge dislocation
is a spiral dislocation, which is a deformation of a circle into a
spiral. The line element describing the space-time backdrop with this
the topological defect is (using the units $\hbar=c=1$) \citep{key-51,key-52,key-53,key-61,key-62,key-63,key-64}
\begin{align}
ds^{2} & =g_{\mu\nu}dx^{\mu}dx^{\nu}\nonumber \\
 & =dt^{2}-d\rho^{2}-\left(\alpha\rho\right)^{2}d\varphi^{2}-\left(dz+Jd\varphi\right)^{2}-\left(dx+\frac{\Phi}{2\pi}d\varphi\right)^{2},\label{eq:17}
\end{align}
where $\chi$ is a constant value relating to the defect's distortion.
By $J=\frac{\left|\overrightarrow{b}\right|}{2\pi}$, the parameter
is also connected to the Burgers vector $\overrightarrow{b}$. Here
$-\infty\le t\le+\infty$, $r\ge0$, $0\le\varphi\le2\pi$, $-\infty\le z\le+\infty$,and
$\alpha\in[0,1[\:$: is the angular parameter that defines the angular
deficit $\delta\varphi=2\pi(1-\alpha)$, which is connected to the
string's linear mass density mu by $\alpha=1-4\mu$(It should be noted
that this metric provides an accurate solution to Einstein's field
equations for $0\le\mu<1/4,$, and that by setting $\varphi^{\prime}=\alpha\varphi$,
it represents a flat conical outer space with angle deficit $\delta\phi=8\pi\mu$).

When the metric and inverse metric tensor components are, respectively,
\begin{equation}
g_{\mu\nu}=\left(\begin{array}{ccccc}
1 & 0 & 0 & 0 & 0\\
0 & -1 & 0 & 0 & 0\\
0 & 0 & -\left\{ \left(\alpha\rho\right)^{2}+\left(\frac{\Phi}{2\pi}\right)^{2}+J^{2}\right\}  & -J & -\frac{\Phi}{2\pi}\\
0 & 0 & -J & -1 & 0\\
0 & 0 & -\frac{\Phi}{2\pi} & 0 & -1
\end{array}\right),\label{eq:18}
\end{equation}
and 
\begin{equation}
g^{\mu\nu}=\left(\begin{array}{ccccc}
1 & 0 & 0 & 0 & 0\\
0 & -1 & 0 & 0 & 0\\
0 & 0 & -\frac{1}{\left(\alpha\rho\right)^{2}} & \frac{J}{\left(\alpha\rho\right)^{2}} & \frac{\Phi}{2\pi\left(\alpha\rho\right)^{2}}\\
0 & 0 & \frac{J}{\left(\alpha\rho\right)^{2}} & -\left(1+\frac{J^{2}}{\left(\alpha\rho\right)^{2}}\right) & -\frac{\Phi J}{2\pi\left(\alpha\rho\right)^{2}}\\
0 & 0 & \frac{\Phi}{2\pi\left(\alpha\rho\right)^{2}} & -\frac{\Phi J}{2\pi\left(\alpha\rho\right)^{2}} & -\left(1+\frac{\Phi^{2}}{\left(2\pi\alpha\rho\right)^{2}}\right)
\end{array}\right)\label{eq:19-1}
\end{equation}
In our case, it is simple to see that $R=0$,
implying that space-time is locally flat (no local gravity), and so
the coupling component is vanishing. The condition $\xi=0$ is known
as minimum coupling. However, in massless theory, $\xi$ equals 1/6.
(in 4 dimensions). The equations of motion are then conformally invariant
in this later instance.

A simple computation yields $\mathcal{Y}=0,$ and we then obtain 
\begin{align*}
\mathcal{T}=\frac{1}{\mathcal{F}} & \left(-\frac{d^{2}}{d\rho^{2}}-\frac{1}{\rho}\frac{d}{d\rho}-\frac{1}{\alpha^{2}\rho^{2}}\frac{d^{2}}{d\varphi^{2}}-\left(1+\frac{J^{2}}{\alpha^{2}\rho^{2}}\right)\frac{d^{2}}{dz^{2}}-\left(1+\frac{\Phi^{2}}{\left(2\pi\alpha\rho\right)^{2}}\right)\frac{d^{2}}{dx^{2}}\right)\frac{1}{\mathcal{F}}\\
 & +\frac{1}{\mathcal{F}}\left(\frac{2J}{\alpha^{2}\rho^{2}}\left(\frac{d}{d\varphi}\frac{d}{dz}\right)+\frac{\Phi}{\pi\alpha^{2}\rho^{2}}\left(\frac{d}{d\varphi}\frac{d}{dx}\right)-\frac{\Phi J}{\pi\alpha^{2}\rho^{2}}\left(\frac{d}{dz}\frac{d}{dx}\right)\right)\frac{1}{\mathcal{F}}
\end{align*}
where 
\begin{equation}
\Phi(t,\rho,\varphi,z,x)=\Phi(\rho)e^{-i\left(Et-l\varphi-Kz-\lambda x\right)},\label{eq:25}
\end{equation}
where $j=0,\pm1,\pm2,..$ are the eigenvalues of the $z$-component
of the angular momentum operator. The KG equation \ref{eq:15} may
be written equivalently to the following two coupled equations 
\begin{align}
\left(\mathcal{N}^{2}+\mathcal{T}\right)\phi_{1}+\left(-\mathcal{N}^{2}+\mathcal{T}\right)\phi_{2} & =2\mathcal{N}E\phi_{1}\nonumber \\
-\left(\mathcal{N}^{2}+\mathcal{T}\right)\phi_{2}-\left(-\mathcal{N}^{2}+\mathcal{T}\right)\phi_{1} & =2\mathcal{N}E\phi_{2},\label{eq:26}
\end{align}
The sum and difference of the two previous equations yield a second
order differential equation for the field $\psi$. As a result, the
the radial equation is as follows: 
\begin{equation}
\left[\frac{d^{2}}{dr^{2}}+\frac{1}{r}\frac{d}{dr}-\frac{\zeta^{2}}{r^{2}}+\kappa\right]\psi\left(r\right)=0,\label{eq:27}
\end{equation}
where we have set 
\begin{align*}
\zeta= & \frac{\sqrt{\left(l-JK\right)^{2}+\frac{\Phi^{2}}{\pi^{2}}\left[\frac{\lambda^{2}}{4}-\frac{l\lambda}{\Phi}+\frac{JK\lambda\pi}{\Phi}\right]}}{\alpha}\\
\kappa= & \sqrt{E^{2}-m^{2}-K^{2}-\lambda^{2}}
\end{align*}
We can observe that Eq. \eqref{eq:27} is a Bessel equation and its
the general solution is defined by \citep{key-56} 
\begin{equation}
\psi\left(r\right)=A\,J_{|\zeta|}\left(\kappa\rho\right)+B\,Y_{|\zeta|}\left(\kappa\rho\right),\label{eq:29}
\end{equation}
where $J_{|\zeta|}\left(\kappa\rho\right)$ and $Y_{|\zeta|}\left(\kappa\rho\right)$
are the Bessel functions of order $\zeta$ and of the first and the
second kind, respectivement. Here $A$ and $B$ are arbitrary constants.
We notice that at the origin when $\zeta=0$, the function $J_{|\zeta|}\left(\kappa\rho\right)\ne0$.
However, $Y_{|\zeta|}\left(\kappa\rho\right)$ is always divergent
at the origin. In this case, we will consider only $J_{|\zeta|}\left(\kappa\rho\right)$
when $\zeta\ne0$. Hence, we write the solution to Eq. \eqref{eq:27}
as follows 
\begin{equation}
\psi\left(\rho\right)=A\,J_{\frac{\left|\sqrt{\left(l-JK\right)^{2}+\frac{\Phi^{2}}{\pi^{2}}\left[\frac{\lambda^{2}}{4}-\frac{l\lambda}{\Phi}+\frac{JK\lambda\pi}{\Phi}\right]}\right|}{\alpha}}\,\,\left(\sqrt{E^{2}-m^{2}-K^{2}-\lambda^{2}}\,\rho\right),\label{eq:30}
\end{equation}
We can now express the whole two-component wavefunction of the spinless
heavy KG particle in the space-time of a cosmic dislocation using
this solution. 
\begin{equation}
\psi\left(t,\rho,\varphi,z,x\right)=\left|\mathcal{C}_{1}\right|\left(\begin{array}{c}
1+\frac{E}{\mathcal{N}}\\
1-\frac{E}{\mathcal{N}}
\end{array}\right)e^{-i\left(Et-l\varphi-Kz-\lambda x\right)}\,J_{\frac{\left|\sqrt{\left(l-JK\right)^{2}+\frac{\Phi^{2}}{\pi^{2}}\left[\frac{\lambda^{2}}{4}-\frac{l\lambda}{\Phi}+\frac{JK\lambda\pi}{\Phi}\right]}\right|}{\alpha}}\,\,\left(\sqrt{E^{2}-m^{2}-K^{2}-\lambda^{2}}\,\rho\right),\label{eq:31}
\end{equation}

The constant $\left|\mathcal{C}_{1}\right|$ can be obtained by applying the appropriate normalization condition to the KG equation (e.g., see Ref.\citep{key-62,key-63}), but it is fortunate that failing to determine the normalization constants throughout this manuscript has no effect on the final results. The wave function that represents the system for values of ($\ell$= 0, K=0), is fully compatible with the energy of all system if study
in \citep{key-45,key-46}.

\subsection{Effects of a hard-wall potential}

From now on, we consider that the scalar particle analyzed in the
previous section is interacting with a hard-wall potential. This type
of potential is characterized by the following boundary condition
\begin{equation}
\psi(\rho_{0})=0,\label{eq01}
\end{equation}
{\color{red} which, in the field of mathematics, is known in the literature as the Dirichlet boundary condition,} and $\rho_{0}$ is a fixed point.

This potential type has been investigated in several systems of quantum
mechanics. For example, on the nonrelativistic oscillator \cite{me},
on Dirac and Klein-Gordon oscillator in global monopole spacetime
\cite{me1}, on relativistic Landau quantization in spacetime with
torsion \cite{me2}, in possible scenarios of Lorentz symmetry violation
\cite{me3,me4} and induced by the noninertial effects in spacetime
with axial symmetry \cite{me5,me6,kb}. Therefore, to complete our
analysis, let us consider the particular case $\sqrt{E^{2}-m^{2}-K^{2}-\lambda^{2}}\rho_{0}>>0$.
In this particular case, the Bessel function is rewritten as follows
\begin{equation}
J_{\frac{\sqrt{\left(l-JK\right)^{2}+\frac{\Phi^{2}}{\pi^{2}}\left[\frac{\lambda^{2}}{4}-\frac{l\lambda}{\Phi}+\frac{JK\lambda\pi}{\Phi}\right]}}{\alpha}}\to\cos{\left[\sqrt{E^{2}-m^{2}-K^{2}-\lambda^{2}}\rho_{0}-\frac{\pi}{4}-\frac{\sqrt{\left(l-JK\right)^{2}+\frac{\Phi^{2}}{\pi^{2}}\left[\frac{\lambda^{2}}{4}-\frac{l\lambda}{\Phi}+\frac{JK\lambda\pi}{\Phi}\right]}}{2\alpha}\right]}\label{eq02}
\end{equation}

By substituting Eq. (\ref{eq02}) into Eq. (\ref{eq:31}), we obtain
from the boundary condition givens in Eq. (\ref{eq01}) the expression
\begin{equation}
E_{K,\lambda,l,n}=\pm\left[m^{2}+K^{2}+\lambda^{2}+\frac{\pi^{2}}{\rho_{0}^{2}}\left(n+\frac{\sqrt{\left(l-JK\right)^{2}+\frac{\Phi^{2}}{\pi^{2}}\left[\frac{\lambda^{2}}{4}-\frac{l\lambda}{\Phi}+\frac{JK\lambda\pi}{\Phi}\right]}}{2\alpha}+\frac{3}{4}\right)^{2}\right]^{1/2}.\label{eq03}
\end{equation}

Eq. (\ref{eq03}) represents the relativistic energy levels of scalar
particle subjected to the hard-wall potential in nontrivial spcetime
composted of curvature and torsion, both characterized by the parameters
$\alpha$ and $J$, which has a extra dimension Kaluza-Klein-type. {\color{red} By making $\lambda \to 0$, $J \to 0$ and $\alpha \to 1$ we obtain the relativistic energy profile of a scalar particle by interacting with a hard-wall potential in Minkowski spacetime, the same particular result obtained in Ref. \citep{me6a}.}

\subsection{Feshbach-Villars oscillator in the background of a cosmic string
space-time in Kaluza--Klein theory}

Now we'll look at the specific instance where we wish to extend the
GFVT for the KGO. In general, we must substitute the momentum operator
in Eq. \eqref{eq:15}. As a result, Eq. \eqref{eq:29} may be rewritten
as follows. 
\begin{align}
\mathcal{T} & =\frac{1}{\mathcal{F}}\left[-\left(\frac{d}{d\rho}-m\omega\rho\right)\left(\sqrt{-g}\right)\left(\frac{d}{d\rho}+m\omega\rho\right)-\left(\sqrt{-g}\right)\left(\frac{1}{\alpha^{2}\rho^{2}}\frac{d^{2}}{d\varphi^{2}}+\left(1+\frac{J^{2}}{\alpha^{2}\rho^{2}}\right)\frac{d^{2}}{dz^{2}}+\left(1+\frac{\Phi^{2}}{\left(2\pi\alpha\rho\right)^{2}}\right)\frac{d^{2}}{dx^{2}}\right)\right]\frac{1}{\mathcal{F}}\nonumber \\
 & +\left(\sqrt{-g}\right)\frac{1}{\mathcal{F}}\left(\frac{2J}{\alpha^{2}\rho^{2}}\left(\frac{d}{d\varphi}\frac{d}{dz}\right)+\frac{\Phi}{\pi\alpha^{2}\rho^{2}}\left(\frac{d}{d\varphi}\frac{d}{dx}\right)-\frac{\Phi J}{\pi\alpha^{2}\rho^{2}}\left(\frac{d}{dz}\frac{d}{dx}\right)\right)\frac{1}{\mathcal{F}}\label{eq:32}
\end{align}
Similarly, the following differential equation may be obtained using
a simple calculation based on the approach described above. 
\begin{equation}
\left[\frac{d^{2}}{d\rho^{2}}+\frac{1}{\rho}\frac{d}{d\rho}+m^{2}\omega^{2}\rho^{2}-\frac{\sigma^{2}}{\rho^{2}}+\delta\right]\psi\left(r\right)=0,\label{eq:33}
\end{equation}
with 
\begin{align}
\sigma^{2}= & \left(\frac{\left(l-JK\right)^{2}+\frac{\Phi^{2}}{\pi^{2}}\left[\frac{\lambda^{2}}{4}-\frac{l\lambda\pi}{\Phi}+\frac{JK\lambda\pi}{\Phi}\right]}{\alpha}\right)^{2}\label{eq:34}\\
\delta= & E^{2}-m^{2}-K^{2}-\lambda^{2}+2m\omega.\nonumber 
\end{align}
The FVO for a spin-0 particle in the (1+4) space-time of KKT is given
by Eq. \eqref{eq:32}. To derive the solution to this problem, we
first, suggest a radial coordinate transformation. 
\begin{equation}
\mathcal{Q}=m\omega r^{2},\label{eq:35}
\end{equation}
subsitutuing the expression for $\chi$ into Eq. \eqref{eq:32}, we
obtain 
\begin{equation}
\left[\frac{d^{2}}{d\mathcal{Q}^{2}}+\frac{1}{\mathcal{Q}}\frac{\partial}{d\mathcal{Q}}-\frac{\sigma^{2}}{4\mathcal{Q}^{2}}+\frac{\delta}{4m\omega\mathcal{Q}}-\frac{1}{4}\right]\psi\left(\chi\right)=0.\label{eq:36}
\end{equation}
So, if we look at the asymptotic behavior of the wave function at
the origin and infinity, and we're looking for regular solutions,
we may assume a solution of the type 
\begin{equation}
\psi\left(\mathcal{Q}\right)=\mathcal{Q}^{\frac{\left|\sigma\right|}{2}}e^{-\frac{\mathcal{Q}}{2}}F\left(\mathcal{Q}\right),\label{eq:37}
\end{equation}
As previously, we can plug this back into Eq. \eqref{eq:35}, and
we get 
\begin{equation}
\mathcal{Q}\frac{d^{2}F\left(\mathcal{Q}\right)}{d\mathcal{Q}^{2}}+\left(|\sigma|+1-\mathcal{Q}\right)\frac{dF\left(\mathcal{Q}\right)}{d\mathcal{Q}}-\left(\frac{|\sigma|}{2}-\frac{\delta}{4m\omega}+\frac{1}{2}\right)F\left(\mathcal{Q}\right)=0,\label{eq:38}
\end{equation}
This is the confluent hypergeometric equation \citep{key-64}, the
solutions which are defined in terms of the kind of confluent hypergeometric
function. 
\begin{equation}
F\left(\mathcal{Q}\right)={}_1 F_{1} \left(\frac{\left|\sigma\right|}{2}-\frac{\delta}{4m\omega}+\frac{1}{2},|\sigma|+1,\mathcal{Q}\right),\label{eq:39}
\end{equation}
We should note that the solution \eqref{eq:38} must be a polynomial
the function of degree $n$. However, taking $n\rightarrow\infty$
imposes a divergence issue. We can have a finite polynomial only if
the factor of the last term in Eq. \eqref{eq:37} is a negative integer,
meaning, 
\begin{equation}
\frac{\left|\sigma\right|}{2}-\frac{\delta}{4m\omega}+\frac{1}{2}=-n\qquad (n=0,1,2,\cdots).\label{eq:40}
\end{equation}
With this result and the parameters \eqref{eq:34}, we may derive
the quantized energy spectrum of FVO in the cosmic dislocation space-time,
and hence, 
\begin{equation}
E^{\pm}\left(n\right)=\pm\sqrt{4m\omega n+\frac{2m\omega}{\alpha}\left|\left(l-JK\right)^{2}+\frac{\Phi^{2}}{\pi^{2}}\left[\frac{\lambda^{2}}{4}-\frac{l\pi\lambda}{\Phi}+\frac{JK\lambda\pi}{\Phi}\right]\right|+m^{2}+K^{2}+\lambda^{2}},\label{eq:41}
\end{equation}
At this stage, some remarks can be make
\begin{itemize}
\item The energy that represents the system for values of ($\ell$ = 0,
K=0 ), is fully compatible with the energy of all system if study
in \citep{key-45,key-46}.

\item The energy that represents the system for values of ($\ell$ = 0,
K=0,$\lambda=0$), is fully compatible with the energy of all system
if study in \citep{key-27}.

\item The energy that represents the system for values of ($\ell$ = 0
, $K=0$, $\lambda=0$ ), and in absence of topological defect is
fully compatible with the energy $E^{\pm}\left(n\right)=\pm\sqrt{4m\omega n+m^{2}}$.
We may notice that the energy relies clearly on the angular deficit
$\alpha$. In other words, because of the presence of the wedge angle,
the curvature of space-time that is impacted by the topological defect,
i.e., the cosmic string would affect the relativistic dynamics of
the scalar particle by creating a gravitational field. The corresponding wave function is given by 
\begin{equation}
\psi\left(\rho\right)=\left|\mathcal{C}_{2}\right|\left(m\omega\rho^{2}\right)^{\frac{\left|\sigma\right|}{2}}e^{-\frac{m\omega\rho^{2}}{2}}{}_{1}F_{1}\left(\frac{\left|\sigma\right|}{2}-\frac{\delta}{4m\omega}+\frac{1}{2},|\sigma|+1,m\omega\rho^{2}\right),\label{eq:42}
\end{equation}
Thereafter, the general eigenfunctions are written as 
\begin{equation}
\psi\left(t,\rho,\varphi,z,x\right)=\left|\mathcal{C}_{2}\right|\left(\begin{array}{c}
1+\frac{E}{\mathcal{N}}\\
1-\frac{E}{\mathcal{N}}
\end{array}\right)\left(m\omega\rho^{2}\right)^{\frac{\left|\sigma\right|}{2}}e^{-\frac{m\omega\sigma^{2}}{2}}e^{-i\left(Et-l\varphi-Kz-\lambda x\right)}{}_{1}F_{1}\left(\frac{\left|\sigma\right|}{2}-\frac{\delta}{4m\omega}+\frac{1}{2},|\sigma|+1,m\omega\rho^{2}\right),\label{eq:43}
\end{equation}
where $\left|\mathcal{C}_{2}\right|$ is the normalization constant
and $\left(\sigma,\delta\right)$are given in \eqref{eq:33}.

\item The wave function that represents the system for values of ( $\ell$
= 0, K=0 ), is fully compatible with the energy of all system if study
in \citep{key-45,key-46}.
\end{itemize}

\subsection{Feshbach-Villars oscillator plus hard-wall potential in the background
of a cosmic string space-time in Kaluza--Klein theory}

From now on, let us consider the relativistic quantum oscillator interacting
with a rigid wall potential defined by the boundary condition given
in Eq. (\ref{eq01}). In order to obtain the relativistic energy profile
of this system, let us consider the particular case $\delta/4m\omega>>1$.
In addition, the quantum number $l$ and $\rho_0$ are fixed. 
{\color{red} In this particular case, the confluent hypergeometric function can be written in
the form \citep{key-56}:
\begin{equation}\label{eq43a}
{}_1 F_1 \left(A,B;\mathcal{Q}_0\right)\to \cos\left(\sqrt{2\,B\,\mathcal{Q}_0-4\,A\mathcal{Q}_0}\right)-\frac{B\pi}{2}+\frac{\pi}{4},
\end{equation}
where we have defined
\begin{equation}\label{eq43b}
A=\frac{\left|\sigma\right|}{2}-\frac{\delta}{4m\omega}+\frac{1}{2}; \ \ \ B=|\sigma|+1.
\end{equation}
Therefore, by substituting Eqs. (\ref{eq43a}) and (\ref{eq:39}) into Eq. (\ref{eq03}) we obtain}
\begin{equation}
E_{K,\lambda,l,n}=\pm\left[m^{2}+K^{2}+\lambda^{2}-2m\omega+\frac{\pi^{2}}{\rho_{0}^{2}}\left(n+\frac{\sqrt{\left(l-JK\right)^{2}+\frac{\Phi^{2}}{\pi^{2}}\left[\frac{\lambda^{2}}{4}-\frac{l\lambda}{\Phi}+\frac{JK\lambda\pi}{\Phi}\right]}}{2\alpha}+\frac{3}{4}\right)^{2}\right]^{1/2}.\label{eq04}
\end{equation}

Eq. (\ref{eq04}) represents the relativistic energy levels of FV
oscillator by interacting with a hard-wall potential in a Kaluza-Klein-type
spacetime with torsion and curvature. By comparing Eqs. (\ref{eq:41})
and (\ref{eq04}), we can note that the presence of the hard-wall
potential changes the relativistic energy profile of FV oscillator.
In addition, by comparing Eqs. (\ref{eq03}) and (\ref{eq04}), we
can observe that the presence of FV oscillator modifies the relativistic
energy levels of hard-wall potential in spacetime defined by the line
element (\ref{eq:17}). This modification is explicit by the term
$-2m\omega$.

\section{FV oscillator in cosmic dislocation in Som--Raychaudhuri spacetime
in Kaluza--Klein theory}

New observational data suggest that the universe is both expanding
and rotating, which has led to increased interest in developing theories
to explain these phenomena. One such theory was proposed by Gödel
in the 1950s for a universe with rigid rotation, characterized by
a term in the metric and a curvature source is known as a Weyssenhoff-Raabe
fluid \citep{key-57,key-58}. Researchers have studied the quantum
dynamics of this theory in (3+1)-dimensional spacetimes \citep{key-59,key-61}.
The authors of this paper focus on cosmic dispiration in a flat Gödel
or Som-Raychaudhuri solution in Kaluza-Klein theory. They assume that
a charged scalar particle is exposed to a uniform magnetic field,
which is introduced using Kaluza-Klein theory through the geometry
of spacetime. Because rotation is important in current scenarios,
the authors use Kaluza-Klein theory to describe the quantum dynamics
of a Klein-Gordon particle in this spinning background. They present
a new solution for the Som-Raychaudhuri spacetime that includes a
topological defect of cosmic dispiration, which is localized parallel
to the rotation axis. The authors consider the influence of introducing
a cosmic string in a Gödel-type universe, in contrast to their previous
section where they studied quantum dynamics in a topological defect
background. Specifically, they consider the Som-Raychaudhuri solution
of the Einstein field equation \citep{key-58} with a cosmic dispiration,
described in a Kaluza-Klein theory with a spinning torsion source
along the symmetry axis of the background spacetimes. 
\begin{equation}
ds^{2}=\left(dt+\alpha\Omega\rho^{2}d\varphi\right)^{2}-d\rho^{2}-\left(\alpha\rho\right)^{2}d\varphi^{2}-\left(dz+Jd\varphi\right)^{2}-\left(dx+\left(\frac{\Phi}{2\pi}+\frac{eB\rho^{2}}{2}\right)d\varphi\right)^{2}\label{eq:44}
\end{equation}
where the components of the metric and the inverse metric tensors
are, respectively, 
\begin{equation}
g_{\mu\nu}=\left(\begin{array}{ccccc}
1 & 0 & \alpha\Omega\rho^{2} & 0 & 0\\
0 & -1 & 0 & 0 & 0\\
\alpha\Omega\rho^{2} & 0 & -\left[\left(\alpha\Omega\rho^{2}\right)^{2}+\left(\frac{\Phi}{2\pi}+\frac{eB\rho^{2}}{2}\right)^{2}+J^{2}\right] & -J & -\left[\frac{\Phi}{2\pi}+\frac{eB\rho^{2}}{2}\right]\\
0 & 0 & -J & -1 & 0\\
0 & 0 & -\left[\frac{\Phi}{2\pi}+\frac{eB\rho^{2}}{2}\right] & 0 & -1
\end{array}\right)\label{eq:45}
\end{equation}
and 
\begin{equation}
g^{\mu\nu}=\left(\begin{array}{ccccc}
1-\left(\Omega\rho\right)^{2} & 0 & \frac{\Omega}{\alpha} & -\frac{\Omega J}{\alpha} & -\left[\frac{\Omega\Phi}{2\pi\alpha}+\frac{eB\Omega}{2\alpha}\rho^{2}\right]\\
0 & -1 & 0 & 0 & 0\\
\frac{\Omega}{\alpha} & 0 & -\frac{1}{\left(\alpha\rho\right)^{2}} & \frac{J}{\left(\alpha\rho\right)^{2}} & \left[\frac{\Phi}{2\pi\left(\alpha\rho\right)^{2}}+\frac{eB}{2\alpha^{2}}\right]\\
0 & 0 & \frac{J}{\left(\alpha\rho\right)^{2}} & -\left(1+\frac{J^{2}}{\left(\alpha\rho\right)^{2}}\right) & -\left[\frac{\Phi J}{2\pi\left(\alpha\rho\right)^{2}}+\frac{eJB}{2\alpha^{2}}\right]\\
-\left[\frac{\Omega\Phi}{2\pi\alpha}+\frac{eB\Omega}{2\alpha}\rho^{2}\right] & 0 & \left[\frac{\Phi}{2\pi\left(\alpha\rho\right)^{2}}+\frac{eB}{2\alpha^{2}}\right] & -\left[\frac{\Phi J}{2\pi\left(\alpha\rho\right)^{2}}+\frac{eJB}{2\alpha^{2}}\right] & -\left[1+\frac{\Phi^{2}}{\left(2\pi\alpha\rho\right)^{2}}+\frac{e\Phi B}{2\pi\alpha^{2}}+\left(\frac{eB}{2\alpha}\rho\right)^{2}\right]
\end{array}\right)\label{eq:46}
\end{equation}

\subsection{Free FV equation in cosmic dislocation in Som-Raychaudhuri spacetime in Kaluza--Klein theory}

In the next part of the mathematical calculations we consider the
magnetic field to be non-existent(B=0),where 
\begin{equation}
g_{\mu\nu}=\left(\begin{array}{ccccc}
1 & 0 & \alpha\Omega\rho^{2} & 0 & 0\\
0 & -1 & 0 & 0 & 0\\
\alpha\Omega\rho^{2} & 0 & -\left[\left(\alpha\Omega\rho^{2}\right)^{2}+\left(\frac{\Phi}{2\pi}\right)^{2}+J^{2}\right] & -J & -\left[\frac{\Phi}{2\pi}\right]\\
0 & 0 & -J & -1 & 0\\
0 & 0 & -\left[\frac{\Phi}{2\pi}\right] & 0 & -1
\end{array}\right)\label{eq:47}
\end{equation}
and 
\begin{equation}
g^{\mu\nu}=\left(\begin{array}{ccccc}
1-\left(\Omega\rho\right)^{2} & 0 & \frac{\Omega}{\alpha} & -\frac{\Omega J}{\alpha} & -\left[\frac{\Omega\Phi}{2\pi\alpha}\right]\\
0 & -1 & 0 & 0 & 0\\
\frac{\Omega}{\alpha} & 0 & -\frac{1}{\left(\alpha\rho\right)^{2}} & \frac{J}{\left(\alpha\rho\right)^{2}} & \left[\frac{\Phi}{2\pi\left(\alpha\rho\right)^{2}}\right]\\
0 & 0 & \frac{J}{\left(\alpha\rho\right)^{2}} & -\left(1+\frac{J^{2}}{\left(\alpha\rho\right)^{2}}\right) & -\left[\frac{\Phi J}{2\pi\left(\alpha\rho\right)^{2}}\right]\\
-\left[\frac{\Omega\Phi}{2\pi\alpha}\right] & 0 & \left[\frac{\Phi}{2\pi\left(\alpha\rho\right)^{2}}\right] & -\left[\frac{\Phi J}{2\pi\left(\alpha\rho\right)^{2}}\right] & -\left[1+\frac{\Phi^{2}}{\left(2\pi\alpha\rho\right)^{2}}\right]
\end{array}\right)\label{eq:48}
\end{equation}
Similarly, the following differential equation may be obtained using
a simple calculation based on the approach described above.

\begin{equation}
\left[\frac{d^{2}}{d\rho^{2}}+\frac{1}{\rho}\frac{d}{d\rho}-\frac{\xi^{2}}{\rho^{2}}+\kappa'\right]\psi\left(r\right)=0\label{eq:49}
\end{equation}
with 
\begin{align}
\xi^{2}= & \left(\frac{\left(l-JK\right)^{2}+\frac{\Phi^{2}}{\pi^{2}}\left[\frac{\lambda^{2}}{4}-\frac{l\pi\lambda}{\Phi}+\frac{JK\lambda\pi}{\Phi}\right]}{\alpha}\right)^{2}\nonumber \\
\kappa'= & E^{2}-m^{2}-K^{2}-\lambda^{2}+\left(\frac{\Omega}{\alpha}\right)^{2}\left[l-JK-\frac{\Phi}{2\pi}\lambda\right]^{2}\label{eq:50}
\end{align}
We can observe that the solution of the Equation is a Bessel equation
and its general solution is defined by 
\begin{equation}
\psi\left(r\right)=A'\,J_{|\zeta|}\left(\kappa'\rho\right)+B'\,Y_{|\zeta|}\left(\kappa'\rho\right)\label{eq:51}
\end{equation}
where $J_{|\xi|}\left(\kappa'\rho\right)$ and $Y_{|\xi|}\left(\kappa'\rho\right)$
are the Bessel functions of order $\xi$ and of the first and the
second kind, respectivement. Here $A'$ and $B'$ are arbitrary constants.
We notice that at the origin when $\xi=0$, the function $J_{|\xi|}\left(\kappa\rho\right)\ne0$.
However, $Y_{|\xi|}\left(\kappa'\rho\right)$ is always divergent
at the origin. In this case, we will consider only $J_{|\xi|}\left(\kappa'\rho\right)$
when $\xi\ne0$. Hence, we write the solution as follows 
\begin{equation}
\psi\left(\rho\right)=A'\,J_{\frac{\left|\sqrt{\left(l-JK\right)^{2}+\frac{\Phi^{2}}{\pi^{2}}\left[\frac{\lambda^{2}}{4}-\frac{l\pi\lambda}{\Phi}+\frac{JK\lambda\pi}{\Phi}\right]}\right|}{\alpha}}\,\,\left(\sqrt{E^{2}-m^{2}-K^{2}-\lambda^{2}+\left(\frac{\Omega}{\alpha}\right)^{2}\left[l-JK-\frac{\Phi}{2\pi}\lambda\right]^{2}}\,\rho\right)\label{eq:52}
\end{equation}
We can now express the whole two-component wavefunction of the spinless
heavy KG particle in the cosmic dislocation in Som--Raychaudhuri
spacetime in Kaluza--Klein theory as 
\begin{equation}
\psi\left(t,\rho,\varphi,z,x\right)=\left|\mathcal{C}_{3}\right|\left(\begin{array}{c}
1+\frac{E}{\mathcal{N}}\\
1-\frac{E}{\mathcal{N}}
\end{array}\right)e^{-i\left(Et-l\varphi-Kz-\lambda x\right)}\,J_{\frac{\left|\sqrt{\left(l-JK\right)^{2}+\frac{\Phi^{2}}{\pi^{2}}\left[\frac{\lambda^{2}}{4}-\frac{l\lambda}{\Phi}+\frac{JK\lambda\pi}{\Phi}\right]}\right|}{\alpha}}\,\,\left(\sqrt{E^{2}-m^{2}-K^{2}-\lambda^{2}}\,\rho\right),\label{eq:53}
\end{equation}

The wave function that represents the system for values of ($\ell$
= 0, K=0 ), is fully compatible with the energy of all system if study
in \citep{key-45,key-46}.

\subsection{Effects of a hard-wall potential}

With the results obtained in the previous subsection, we can go further in order to obtain the relativistic energy levels through the presence of a rigid wall potential, given in Eq. (\ref{eq01}). Therefore,
following the same steps between Eqs. (\ref{eq01}) and (\ref{eq03}),
we obtain the following expression: 
\begin{equation}
E_{K,\lambda,l,n}=\pm\left[m^{2}+K^{2}+\lambda^{2}-\left(\frac{\Omega}{\alpha}\right)^{2}\left(l-Jk-\frac{\Phi\lambda}{2\pi}\right)^{2}+\frac{\pi^{2}}{\rho_{0}^{2}}\left(n+\frac{\sqrt{\left(l-JK\right)^{2}+\frac{\Phi^{2}}{\pi^{2}}\left[\frac{\lambda^{2}}{4}-\frac{l\lambda}{\Phi}+\frac{JK\lambda\pi}{\Phi}\right]}}{2\alpha}+\frac{3}{4}\right)^{2}\right]^{1/2},\label{eq05}
\end{equation}
which represents the relativistic energy levels of a scalar particle
subjected to a hard-wall potential in Som-Raychaudhuri spacetime with
a Kaluza-Klein-type extra dimension. We can note that the system is
influenced by the non trivial spacetime topology. By compraing the
results given Eq. (\ref{eq03}) and (\ref{eq05}), we can note that
the intrinsic gravitational effects of the Som-Raychaudhuri metric
modifies the energy levels of the scalar particle interacting with
the hard-wall potential through the term $\frac{\Omega}{\alpha}$;
by taking $\Omega\to0$, we recover the result obtained in Eq. (\ref{eq03}).

\subsection{FV oscillator in cosmic dislocation in Som--Raychaudhuri spacetime
in Kaluza--Klein theory}

Similarly, the following differential equation may be obtained using
a simple calculation based on the approach described above. 
\begin{equation}
\left[\frac{d^{2}}{d\rho^{2}}+\frac{1}{\rho}\frac{d}{d\rho}+m^{2}\omega^{2}\rho^{2}-\frac{\mu^{2}}{\rho^{2}}+\eta\right]\psi\left(r\right)=0\label{eq:54}
\end{equation}
with 
\begin{align}
\mu^{2}= & \left(\frac{\left(l-JK\right)^{2}+\frac{\Phi^{2}}{\pi^{2}}\left[\frac{\lambda^{2}}{4}-\frac{l\lambda\pi}{\Phi}+\frac{JK\lambda\pi}{\Phi}\right]}{\alpha}\right)^{2}\nonumber \\
\eta= & E^{2}-m^{2}-K^{2}-\lambda^{2}+\left(\frac{\Omega}{\alpha}\right)^{2}\left[l-JK-\frac{\Phi}{2\pi}\lambda\right]^{2}+2m\omega\label{eq:55}
\end{align}
With this result and the parameters \eqref{eq:34}, we may derive
the quantized energy spectrum of FVO in the cosmic dislocation space-time,
and hence, 
\begin{equation}
E^{\pm}\left(n\right)=\pm\sqrt{4m\omega n+\frac{2m\omega}{\alpha}\left|\left(l-JK\right)^{2}+\frac{\Phi^{2}}{\pi^{2}}\left[\frac{\lambda^{2}}{4}-\frac{l\lambda\pi}{\Phi}+\frac{JK\lambda\pi}{\Phi}\right]\right|+\left(\frac{\Omega}{\alpha}\right)^{2}\left[l-JK-\frac{\Phi}{2\pi}\lambda\right]^{2}+m^{2}+K^{2}+\lambda^{2}}\label{eq:56}
\end{equation}
Some remarks can be made here
\begin{itemize}
\item The energy that represents the system for values of ($\ell$ = 0, K=0 ), is fully compatible with the energy of all system if study in \citep{key-45,key-46}.
\item The energy that represents the system for values of ($\ell$ = 0,K=0,$\lambda=0$), is fully compatible with the energy of all system if study in \citep{key-27}.
\item The energy that represents the system for values of ($\ell$ = 0, $K=0$, $\lambda=0$ ), and in absence of topological defect is
fully compatible with the energy $E^{\pm}\left(n\right)=\pm\sqrt{4m\omega n+m^{2}}$.
We may notice that the energy relies clearly on the angular deficit
$\alpha$. In other words, because of the presence of the wedge angle,
the curvature of space-time that is impacted by the topological defect,
i.e., the cosmic string would affect the relativistic dynamics of
the scalar particle by creating a gravitational field.

The corresponding wave function is given by 
\begin{equation}
\psi\left(\rho\right)=\left|\mathcal{C}_{4}\right|\left(m\omega\rho^{2}\right)^{\frac{\left|\mu\right|}{2}}e^{-\frac{m\omega\rho^{2}}{2}}{}_{1}F_{1}\left(\frac{\left|\mu\right|}{2}-\frac{\eta}{4m\omega}+\frac{1}{2},|\mu|+1,m\omega\rho^{2}\right)\label{eq:57}
\end{equation}
Thereafter, the general eigenfunctions are written as 
\begin{equation}
\psi\left(t,\rho,\varphi,z,x\right)=\left|\mathcal{C}_{4}\right|\left(\begin{array}{c}
1+\frac{E}{\mathcal{N}}\\
1-\frac{E}{\mathcal{N}}
\end{array}\right)\left(m\omega\rho^{2}\right)^{\frac{\left|\mu\right|}{2}}e^{-\frac{m\omega\rho^{2}}{2}}e^{-i\left(Et-l\varphi-Kz-\lambda x\right)}{}_{1}F_{1}\left(\frac{\left|\mu\right|}{2}-\frac{\eta}{4m\omega}+\frac{1}{2},|\mu|+1,m\omega\rho^{2}\right)\label{eq:58}
\end{equation}
where $\left|\mathcal{C}_{4}\right|$ is the normalization constant
and $\left(\mu,\eta\right)$are given in \eqref{eq:54}.
\item The wave function that represents the system for values of ($\ell$
= 0, K=0 ), is fully compatible with the energy of all system if study
in \citep{key-45,key-46}.
\end{itemize}

\subsection{FV oscillator plus a hard-wall potential in cosmic dislocation in Som--Raychaudhuri spacetime in Kaluza--Klein theory}

Let us consider the FV oscillator by interacting with a hard-wall
potential defined by the boundary condition given in Eq. (\ref{eq01}).
By analogy, let us follow the same steps given in Eqs. (\ref{eq01})
to (\ref{eq03}), such that we have 
\begin{eqnarray}
E_{K,\lambda,l,n}&=&\pm\Bigg[m^{2}+K^{2}+\lambda^{2}-2m\omega-\left(\frac{\Omega}{\alpha}\right)^{2}\left(l-Jk-\frac{\Phi\lambda}{2\pi}\right)^{2}\nonumber\\
&&+\frac{\pi^{2}}{\rho_{0}^{2}}\left(n+\frac{\sqrt{\left(l-JK\right)^{2}+\frac{\Phi^{2}}{\pi^{2}}\left[\frac{\lambda^{2}}{4}-\frac{l\lambda}{\Phi}+\frac{JK\lambda\pi}{\Phi}\right]}}{2\alpha}+\frac{3}{4}\right)^{2}\Bigg]^{\frac{1}{2}}.\label{eq06}
\end{eqnarray}

Eq. (\ref{eq06}) give us the relativistic energy levels of FV oscillator
by interacting with a hard-wall potential in Som-Raychaudhuri spacetime
with a Kaluza-Klein-type extra dimension. By comparing Eqs. (\ref{eq05})
and (\ref{eq06}), we can note that the presence of FV oscillator
modifies the relativistic energy spectrum of this quantum system.
This modification comes through of the term $-2m\omega$. By making
$\omega\to0$ in Eq. (\ref{eq06}) we recover the result into Eq.
(\ref{eq05}); by making $\Omega\to0$ into Eq. (\ref{eq05}) we recover
Eq. (\ref{eq03}).

\section{Conclusions and summary}

{\color{red} 

We are currently investigating the quantum dynamics of a relativistic quantum oscillator called the FV oscillator in a space-time that exhibits both torsion and curvature. In this space-time, there is a presence of a screw dislocation and a cosmic string, and we have introduced a quantum flux through an additional Kaluza-Klein-type extra dimension. Initially, we proposed to study a simpler scenario by considering a scalar particle in this non-trivial space-time. We have successfully determined the general solutions for this scalar particle in terms of the first type Bessel function. These solutions are influenced by the background geometry, specifically the torsion, curvature, and extra dimension. The dependence on these background parameters is observed through the eigenfunctions of the system.

In addition to studying the scalar particle in the non-trivial space-time with torsion, curvature, and an extra dimension, we have introduced a hard-wall potential by imposing a boundary condition. By considering a particular case, we have successfully determined the relativistic energy levels of this system. These energy levels are defined in terms of the parameters associated with the background, such as the torsion, curvature, and the extra dimension. The profile of the system's energy levels depends on these background parameters, and an interesting phenomenon similar to the Aharanov-Bohm effect for bound states emerges. Specifically, the parameters associated with the screw displacement, cosmic string, and extra dimension give rise to an effective angular momentum. This effective angular momentum influences the energy levels of the system and contributes to their dependence on the background parameters.

In our investigation, we extend the analysis by incorporating the FV-oscillator into the wave equation through a non-minimal coupling. This allows us to explore the gravitational effects on this relativistic quantum oscillator model. By performing analytical analysis, we determine the energy levels of the system, which are influenced by the presence of torsion, curvature, and the extra dimension. The dependency of the energy levels on the background parameters, namely the torsion, curvature, and the extra dimension, is explicitly expressed in our findings. Specifically, the energy profile of the system is affected by the parameters associated with the screw dislocation, cosmic string, and the quantum flux originating from the extra dimension.

Continuing our investigation, we incorporate the presence of a hard-wall potential that interacts with the FV oscillator, characterized by a specific boundary condition. Through our analysis, we observe that this interaction leads to significant modifications in the energy levels of the system. However, these energy levels still retain their dependence on the parameters associated with torsion, curvature, and the extra dimension. Despite the substantial modifications caused by the interaction with the hard-wall potential, the underlying influence of the background parameters, including torsion, curvature, and the extra dimension, remains evident in the energy profile of the system. 

We extend our analysis to investigate the FV-oscillator in a G\"{o}del-type space-time, specifically the Som-Raychaudhuri space-time in the Kaluza-Klein theory. This space-time possesses an intrinsic curvature, a cosmic string, a screw dislocation, and a Kaluza-Klein-type extra dimension. By considering this background, we repeat all the previous analyses conducted. Remarkably, we observed a generalization of the obtained results, which is evident in the dependence of the energy levels on the vorticity parameter associated with the G\"{o}del-type space-time. This vorticity parameter encapsulates the unique characteristics of the particular Gödel-type background. Thus, we find that the energy levels are influenced by the vorticity, indicating the broader applicability of our findings. To further validate this generalization, we demonstrate that as we approach a vanishing vorticity parameter, our results converge to those obtained in the background featuring only torsion, curvature, and the extra dimension. It is noteworthy that this type of relativistic quantum oscillator can also investigate in other possible scenarios, such as rotational frame \cite{me5}, Lorentz symmetry violation \cite{me7}, central potential \citep{me11}, and external fields \cite{me2} effects. The results obtained in this study can serve for a future thermodynamic analysis.

In summary, this research contributes to the understanding of the relativistic quantum dynamics of spin-0 scalar massive charged particles by investigating their behavior in the context of the Kaluza-Klein Theory. Through the study of the Feshbach-Villars equation and oscillator, as well as their interactions with cosmic string space-time and cosmic dislocations, we gained valuable insights into the quantification of energy, wave function, and the influence of external fields on these systems.
}

\section*{Data Availability Statement}

No new data generated or analyzed in this study.

\section*{Conflict of Interests}

Author's declare(s) no conflict of interests.

\section*{Funding Statement}

No fund has received for this manuscript.

\end{document}